\documentclass[12pt]{article}

\usepackage{amsmath}
\usepackage{amssymb}
\usepackage{amsfonts}
\usepackage{lscape}

\begin{document}

\fontsize{12}{6mm}\selectfont
\setlength{\baselineskip}{2em}

$~$\\[.35in]
\newcommand{\dss}{\displaystyle}
\newcommand{\raro}{\rightarrow}
\newcommand{\be}{\begin{equation}}

\newtheorem{theorem}{Theorem}[section]

\def\sech{\mbox{\rm sech}}
\def\sn{\mbox{\rm sn}}
\def\dn{\mbox{\rm dn}}
\thispagestyle{empty}

\begin{center}
{\Large\bf Exterior Differential Systems,}  \\    [2mm]
{\Large\bf Prolongations and the Integrability of Two}  \\    [2mm]
{\Large\bf Nonlinear Partial Differential Equations}   \\   [2mm]
\end{center}

\vspace{1cm}
\begin{center}
{\bf Paul Bracken}                        \\
{\bf Department of Mathematics,} \\
{\bf University of Texas,} \\
{\bf Edinburg, TX  }  \\
{78541-2999}
\end{center}

\vspace{3cm}
\begin{abstract}
A generalized KdV equation is formulated as an exterior differential
system, which is used to determine the prolongation structure
of the equation. The prolongation structure is obtained  for
several cases of the variable powers, and nontrivial algebras
are determined. The analysis is extended to a differential system 
which gives the Camassa-Holm equation as a particular case.
The subject of conservation laws is briefly discussed for
each of the equations. A B\"acklund transformation is determined
using one of the prolongations.
\end{abstract}

\vspace{2mm}
MSc: 35G20, 35Q53, 37J35, 37K25, 37K35, 55R10, 53Z05.
\vspace{2mm}

\newpage
\section{Introduction.}

Once nonlinear terms are included in linear dispersive
equations, solitary waves can result which can be stable
enough to persist indefinitely.
It is well known that many important nonlinear evolution
equations which have numerous applications
in mathematical physics appear as sufficient conditions
for the integrability of systems of linear partial
differential equations of first order, and such systems 
are referred to as integrable {\bf [1-2]}. This is not
just an oddity, since algebraic structures such as those
which appear in AKNS systems can arise very naturally
from nonlinear evolution equations. This is very well
exemplified by applying the so called prolongation
technique. Wahlquist and Estabrook {\bf [3-5]} first
constructed prolongations, or $sl (2)$ systems, both for the
KdV and nonlinear Schr\"odinger equations, and Shadwick the
former {\bf [6]}. This procedure produces
a nonclosed Lie algebra of vector fields  which are
defined on fibres above the base manifold that supports
the exterior differential forms defining the nonlinear
evolution equation. It has been shown that a simple 
linear prolongation of the KdV, sine-Gordon and
non-linear Schr\"odinger equation can be provided
by $sl (2, \mathbb C)$ {\bf [7]}. What is more, it has been shown
that the vanishing of the curvature form of a
particular Cartan-Ehresmann connection is the necessary
and sufficient condition for the existence of the
prolongation.

These prolongations have a very useful application since
B\"acklund transformations can be calculated based on them as well {\bf [8]}.
A B\"acklund transformation has important practical consequences,
since such transformations can be used to calculate
solutions to an associated equation, usually referred to
as the potential equation, based on solutions of the
initial equation. Sometimes these transformations can be used to obtain new
solutions to the same initial equation, in which case
they are referred to as auto-B\"acklund transformations.

Recently an exterior differential system which defines
a generalized KdV equation on the transverse manifold
was obtained {\bf [9]}. A particular case of this equation
has appeared in {\bf [10]} recently. The symmetries 
of this equation were determined and some solutions 
were found as well {\bf [11]}. This permitted the
determination of a certain form of integrability. Also,
a particular type of prolongation over a fibre bundle was
found corresponding to this differential system, as well as
a specific form for a B\"acklund transformation with its
associated potential equation. Here, the same differential
system is studied, but a fully general calculation of
the prolongation over the same bundle is carried out
in detail for this generalized KdV equation. This allows 
the prolongation structure for any case of the given
parameters in the equation.
For completeness, the general theory 
for obtaining such prolongations based on the given  
exterior system of differential forms that defines the equation
upon sectioning to a transversal integral manifold will be outlined first. 
Transversal integral manifolds give solutions of the equation.
Finally this work is extended 
to a study of a differential system of one-forms which define an equation
that includes the Camassa-Holm equation and Degasperis-Procesi equations
as specific cases {\bf [12-14]}. The Camassa-Holm equation has been of interest
because it has been shown to have peaked soliton solutions. 
The Camassa-Holm equation has alot in common with the KdV equation,
but there are significant differences as well.
The KdV equation is globally well-posed when considered on
a suitable Sobolev space, while Camassa-Holm is in general not.
The first derivative of a solution of the latter can become
infinite in finite time.
The associated
prolongation equations are developed and found to be much more restrictive
than the previous case. However, it is shown that at
least one solution to the prolongation system can be found.
Finally, for each system a brief discussion concerning how
conservation laws arise and can be expressed in this context will be discussed
based on the defining exterior differential system.

\section{Introduction to Cartan Prolongations.}

Cartan prolongations will be found for the equations
mentioned in the Introduction, and to begin, a general
outline of this subject is given. Consider the space
$M = \mathbb R^n$ with the coordinates 
$( x, t, u, p, q, \cdots)$ and let there be given
on $M$ a closed exterior differential system
$$
\alpha_1 =0, \qquad
\cdots,  \qquad
\alpha_l =0. 
\eqno(2.1)
$$
Let $I$ be the ideal generated by the system
(2.1), and so $I = \{ \chi = \sum_{i=1}^l
\sigma_i \wedge \alpha_i : \sigma_i \in 
\Lambda_p (M), p=0,1,2, \cdots \}$ with
$\Lambda_p (M)$ the set of $p$-forms on $M$.
Since (2.1) is closed, we have $d I \subset I$
and (2.1) is integrable.

The system (2.1) will be chosen such that
solutions $u= u(x,t)$ of an evolution equation
$u_t = F (x,t, u, u_x, u_{xx}, \cdots )$
correspond with two-dimensional transversal
integral manifolds of (2.1). These integral
manifolds can be written as sections $S$ in
$M$ with $S$ given by
$$
(x,t) \rightarrow ( x, t, u(x,t), p(x,t), q(x,t),
\cdots ),
$$
and due to transversality, $dx \wedge dt |_S
= \pi^* ( dx \wedge dt) \neq 0$ such that
$\pi :M \rightarrow \mathbb R^2$ and
$\pi^* : \Lambda ( \mathbb R^2) \rightarrow
\Lambda (S)$. Now introduce the fibre bundle
$( \tilde{M}, \tilde{p}, M)$ over $M$, with
$M \subset \tilde{M}$ and $\tilde{p}$ the projection of
$\tilde{M}$ onto $M$, $\tilde{p} ( \tilde{M}) =M$.
Points in $\tilde{M}$ are written $\tilde{m}$, 
those in $M$ written $m$ and $\tilde{p} ( \tilde{m}) =m$. 
The tangent and cotangent spaces of $\tilde{M}$
and $M$ are denoted by $T (\tilde{M} )$ and
$T^* (\tilde{M})$.

For reference, a Cartan-Ehresmann connection in the
fibre bundle $( \tilde{M}, \tilde{p}, M)$ is a
system of 1-forms $\tilde{\omega}_i$, $i=1,2,\cdots, k$
in $ T^* ( \tilde{M})$ with the property that the
mapping $\tilde{p}_*$ from the vector space
$H_{\tilde{m}} = \{ \tilde{X} \in T_{\tilde{m}} |
\tilde{\omega}_i ( \tilde{X} ) =0, i=1,2,\cdots, k \}$
onto the tangent space $T_m$ is a bijection for all
$\tilde{m} \in \tilde{M}$.
At this point, consider the exterior differential
system in $\tilde{M}$
$$
\tilde{\alpha}_i = \tilde{p}^* \alpha_i =0,
\quad
i=1, \cdots, l,
\quad
\tilde{\omega}_i =0,
\quad
i=1, \cdots, k,
\eqno(2.2)
$$
with $\{ \tilde{\omega}_i \}$ a Cartan-Ehresmann
connection in $(\tilde{M}, \tilde{p}, M)$.
The system (2.2) is called a Cartan prolongation
if (2.2) is closed and whenever $S$ is a
transversal solution of (2.1), then there should
also exist a transversal solution $\tilde{S}$
of (2.2) with $\tilde{p} ( \tilde{S} ) =S$.
It follows from (2.2) closed that this prolongation
condition may be written as
$$
d \tilde{\omega}_i = \sum_{j=1}^k \,
\tilde{\beta}^j_i \wedge \tilde{\omega}_j,
\quad
mod \quad \tilde{p}^* (I),
\eqno(2.3)
$$
where $I$ is the ideal defined in (2.1).

For the considerations here, the fibre bundle will
be the trivial fibre bundle given by $\tilde{M}
= M \times \mathbb R^k$ with $y = (y_1, \cdots, y_k)
\in \mathbb R^k$ and the connection used will have the form
$$
\tilde{\omega}_i = d y_i - \eta_i,
\qquad
\eta_i = \, A_i \, dx + B_i \, dt.
\eqno(2.4)
$$
In (2.4), $A_i$ and $B_i$ are defined as 
$C^{\infty}$-functions on $\tilde{M}$, $i=1, \cdots, k$.
The prolongation condition (2.3) applied to (2.4) is then
$$
- d \eta_i = \tilde{\beta}_i^j \wedge
( d y_j \wedge \eta_j),
\quad mod \quad \tilde{p}^* (I),
\quad i=1, \cdots, k,
\eqno(2.5)
$$
with $\eta_i$ given in (2.4) and $A_i$, $B_i$ depend
on $x$, $t$, $u$, $p$, $q, \cdots, y_1, \cdots, y_k$.
Comparing both sides of (2.5),
$\tilde{\beta}^j_i$ cannot contain differentials of
the form $d y_s - \eta_s$ for $s \neq j$ and so
$$
\tilde{\beta}_i^j = a^j_i \, dx + \beta_i^j \, dt
+ c^j_i \, du + d^j_i \, dp + \cdots \quad
mod \quad \gamma_i^j ( dy_j - \eta_j ).
$$
Comparing forms on both sides of (2.5) yields the results
$$
a_i^j = \frac{\partial A_i}{\partial y_j},
\qquad
b_i^j = \frac{\partial B_i}{\partial y_i},
$$
with $c_i^j = d_i^j = \cdots =0$ because
$du \wedge d y_j$, $dp \wedge d y_j, \cdots$ do not
occur on the left of (2.5). The prolongation
condition reduces to
$$
- d \eta_i = \frac{\partial \eta_i}{\partial y_j}
\wedge ( d y_j - \eta_j ),
\quad mod \quad \tilde{p}^* (I).
\eqno(2.6)
$$
Introducing the vertical valued one-form
$\eta = \eta_i \frac{\partial}{\partial y_i}$
as well as the definitions
$$
d \eta = ( d_M \eta_i ) \frac{\partial}{\partial y_i},
\quad
[ \eta, \omega] = ( \eta_j \wedge \frac{\partial \omega_i}
{\partial y_j} + \omega_j \wedge \frac{\partial \eta_i}
{\partial y_j} ) \frac{\partial}{\partial y_i},
\eqno(2.7)
$$
the prolongation condition reduces to the compact
form,
$$
d \eta + \frac{1}{2} [ \eta, \eta ] =0,
\quad mod \quad \tilde{p}^* (I).
\eqno(2.8)
$$
The form on the left of (2.8) is called the
curvature form of the Cartan-Ehresmann connection
$(d y_i - \eta_i)_{i=1}^{\infty}$. Thus, a
sufficient condition for the existence of a
Cartan prolongation of the set of exterior
differential forms (2.1) is the vanishing of the
curvature form of the Cartan-Ehresmann
connection $(dy_i - \eta_i)_{i=1}^k$.

\section{ Cartan Prolongation of a Generalized KdV Equation.}

\subsection{Differential System and Associated Partial Differential Equation.}

Let us introduce the exterior differential system
defined over a base manifold $M = \mathbb R^5$ which supports the differential
forms. Consider the system of two forms given by
$$
\alpha_1 = n u^{n-1} \, du \wedge dt 
- p \, dx \wedge dt =0,
$$
$$
\alpha_2 = dp \wedge dt - q \, dx \wedge dt =0,
\eqno(3.1)
$$
$$
\alpha_3 = du \wedge dt - d q \wedge dt
- \gamma p u^s \, dx \wedge dt =0,
$$
where $\gamma$ is a nonzero, real constant.
The exterior derivatives of the $\alpha_j$ are given
by
$$
d \alpha_1 = - dp \wedge dx \wedge dt = dx 
\wedge \alpha_2,
$$
$$
d \alpha_2 =- dq \wedge dx \wedge dt =- dx \wedge \alpha_3,
\eqno(3.2)
$$
$$
d \alpha_3 =- \gamma sp u^{s-1} \, du \wedge dx \wedge dt
- \gamma u^s dp \wedge dx \wedge dt
= dx \wedge ( \gamma \frac{s}{n} p u^{s-n} \alpha_1
+ \gamma p u^s \alpha_2 ).
$$
Therefore, the ideal $I = \{ \omega | \omega = 
\sum_{i=1}^3 \, \sigma_i \wedge \alpha_i :
\sigma_i \in \Lambda (M) \}$ is closed,
$d I \subset I$ and the system $\{ \alpha_i \}$ given by
(3.1) is integrable. On the transversal integral 
manifold, it follows that differential system (3.1)
can be sectioned to give,
$$
0 = \alpha_1 |_S = S^* \alpha_1 = (( u^n )_x -p) \, dx 
\wedge dt,
$$
$$
0 = \alpha_2 |_S = S^* \alpha_2 = ( p_x -q) \, dx 
\wedge dt,
\eqno(3.3)
$$
$$
0 = \alpha_3 |_S = S^* \alpha_3 = 
( u_t dt \wedge dx - q_x \, dx \wedge dt
- \gamma p u^s \, dx \wedge dt ).
$$
The transversal integral manifolds correspond to the
equations
$$
p = (u^n )_x,
\qquad
q= p_x = ( u^n )_{xx},
\qquad
u_t + q_x + \gamma p u^s =0.
\eqno(3.4)
$$
Suppose that $n+s \neq 0$, then upon
substituting $p$ and $q$ from the first two
equations in (3.4) into the third, it can be
seen that $u$ must satisfy the following
generalized KdV equation
$$
u_t + (u^n )_{xxx} + \gamma \frac{n}{n+s}
( u^{n+s} )_x =0.
\eqno(3.5)
$$
A more compact form is obtained if we set $m= n+s$
and define a new constant $\beta = n \gamma/
(n+s)$ so that (3.5) takes the form
$$
u_t + ( u^n )_{xxx} + \beta ( u^m )_x =0.
\eqno(3.6)
$$
This is the partial differential equation defined
by differential system (3.1) which was studied in {\bf [11]}.

\subsection{Prolongations.}

Based on the forms
in system (3.1), the prolongation method outlined
in Section 2 can be carried out, and the
resulting system of equations can be solved quite
generally. A very general prolongation corresponding
to (3.6) can be calculated in terms of an algebra
of vector fields which are defined on fibres above the 
base manifold that supports the forms (3.1).
To do this, introduce the pseudopotentials and the
Cartan-Ehresmann connection on the trivial fibre
bundle $\tilde{M} = M \times \mathbb R^k = \mathbb R^5
\times \mathbb R^k$ with coordinates $y = (y_1, \cdots,
y_k)$ on $\mathbb R^k$. The connection forms are
taken to be
$$
\tilde{\omega}_i = d y_i - \eta_i,
\qquad
\eta_i = A_i \, dx + B_i \, dt,
\qquad
A_i = A_i (x,t,u,p,q,y),
\qquad
B_i = B_i (x,t,u,p,q,y).
\eqno(3.7)
$$
Substituting $\eta = \eta_i \, \partial/ \partial y_i$,
$A = A_i$ $\partial / \partial y_i$, $B = B_i \partial /
\partial y_i$, the prolongation condition for the
vectors $A$ and $B$ is
$$
( d_M A_i \wedge dx ) \frac{\partial}{\partial y_i} 
+ ( d_M B_i \wedge dt ) \frac{\partial}{\partial y_i}
+ \frac{1}{2}  \{ A_j \frac{\partial B_i}{\partial y_j}
\, dx \wedge dt + B_j \frac{\partial A_i}{\partial y_j}
\, dt \wedge dx \} \frac{\partial}{\partial y_i}
$$
$$
+ \frac{1}{2} \{ B_j \frac{\partial A_i}{\partial y_j} \,
dt \wedge dx + A_j \frac{\partial B_i}{\partial y_j} \,
dx \wedge dt \} \frac{\partial}{\partial y_i}
$$
$$
=  d A \wedge dx + dB \wedge dt + [ A, B] \, dx \wedge dt =0,
\qquad mod \quad \tilde{p}^* (I),
$$
where $[A, B]$ denotes the ordinary Lie bracket of the
vector fields $A_i \partial / \partial y_i$ and
$B_i \partial /\partial y_i$ defined along fibres of the bundle.
Using (3.1), it is found that the prolongation
condition takes the form
$$
\frac{\partial A}{\partial t} \, dt \wedge dx 
+ \frac{\partial A}{\partial u} \, du \wedge dx
+ \frac{\partial A}{\partial p} \, dp \wedge dx
+ \frac{\partial A}{\partial q} \, dq \wedge dx
$$
$$
+ \frac{\partial B}{\partial x} \, dx \wedge dt
+ \frac{\partial B}{\partial u} \, du \wedge dt
+ \frac{\partial B}{\partial p} \, dp \wedge dt
+ \frac{\partial B}{\partial q} \, dq \wedge dt +[A, B] \, dx \wedge dt
$$
$$
= \lambda_1 ( n u^{n-1} du \wedge dt - p \, dx \wedge dt)
+ \lambda_2 ( dp \wedge dt - q \, dx \wedge dt)
+ \lambda_3 ( du \wedge dx - dq \wedge dt -
\gamma p u^s \, dx \wedge dt).
$$
Comparison of both sides of this equation yields the
following set of conditions,
$$
\begin{array}{ccc}
A_u = \lambda_3, & A_p =0,  &  A_q =0,  \\
   &  &   \\
B_u = n \lambda_1 u^{n-1},  &  B_p = \lambda_2, &  B_q =- \lambda_3,  \\
   &  &   \\
& - A_t + B_x + [A,B] =- p \lambda_1 -q \lambda_2 - \gamma p u^s  
\lambda_3.  &   \\
\end{array}
\eqno(3.8)
$$
Subscripts indicate partial differentiation with respect to the
variable indicated.
Translations in $x$ and $t$ constitute symmetries of 
equation (3.6) {\bf [11]}, and so a simplifying 
assumption would be to suppose that $A$ and $B$ are independent
of $x$ and $t$. Then it must be that $A$ and $B$ are also
invariant under translations in these variables.
The prolongation equations to be solved from (3.8) reduce to the following
$$
A_p = 0 , \qquad A_q =0,
\qquad
A_u =- B_q,
\qquad
\frac{1}{n} u^{1-n}p B_u + q B_p - \gamma p u^s B_q =- [A, B].
\eqno(3.9)
$$

\begin{theorem} 
System (3.9) can be reduced to a single
expression which specifies the algebra of brackets of a set of
basis vector fields $X_i$. The structure of these algebras
is dependent on the relative values of $m$ and $n$.
\end{theorem}

{\em Proof:}
The first three differential equations in (3.9) imply the
following results
$$
A= A(u,y), 
\qquad
B = B (u,p,q,y),
\quad
B =- q \, A_u (u,y) + \hat{B} (u,p,y).
\eqno(3.10)
$$
Substituting $B$ from (3.10) into (3.9) and collecting
terms in $q$ gives
$$
q ( - \frac{1}{n} u^{-n+1} p A_{uu}
+ \hat{B}_p - [ A, A_u] ) + \frac{1}{n} p u^{-n+1} \hat{B}_u 
+ \gamma p u^s A_u + [ A, \hat{B} ] =0.
\eqno(3.11)
$$
Since $A$ and $\hat{B}$ do not depend on $q$, it
follows from (3.11) that
$$
\hat{B}_p = \frac{1}{n} u^{-n+1} p A_{uu} +
[ A, A_u ].
$$
As $A$ does not depend on $p$, this can be integrated
to give $\hat{B}$,
$$
\hat{B} (u,p,y) = \frac{1}{2n} u^{-n+1} p^2 A_{uu}
+ [ A, A_u ] p + B'' (u,y).
\eqno(3.12)
$$
Substituting (3.12) into (3.11) as well as $\hat{B}_u$, there results
$$
\frac{1}{2n} u^{-2n+1} ( - (n-1) A_{uu} + u A_{uuu} ) p^3
+ u^{-n+1} [ A, A_{uu} ] p^2 + u^{-n+1} B_u '' p + n \gamma p u^s A_u
$$
$$
+ n [ A, \frac{1}{2n} u^{-n+1} A_{uu} p^2 + [ A, A_u] p + B'' ] =0.
\eqno(3.13)
$$
Since $A$ and $B''$ do not depend on $p$,
the coefficient of $p^3$ must vanish giving the equation
$$
u A_{uuu} - (n-1) A_{uu} =0.
$$
This can be solved for $A$ to give
$$
A ( u,y) = X_1 (y) + X_2 (y) u + X_3 (y) u^{n+1},
\eqno(3.14)
$$
where the $X_i (y)$ are vertical vector fields. Consequently,
(3.13) simplifies to
$$
u^{-n+1} ( [ A, A_{uu} ] + \frac{1}{2} [ A, A_{uu} ]) p^2
+ (n \gamma u^s A_u + u^{-n+1} B_u '' + n [ A, [A, A_u]]) p
+ n [ A, B'' ] =0. 
\eqno(3.15)
$$
The coefficient of $p^2$ implies that $[ A, A_{uu} ]=0$,
which using (3.14) immediately establishes two basic
commutators of the vector fields $X_1$, $X_2$, and $X_3$,
$$
[ X_1 , X_3 ] =0,
\qquad
[ X_2, X_3 ] =0.
\eqno(3.16)
$$
The coefficient of $p$ implies the condition,
$$
n \gamma u^s A_u + u^{-n+1} B_u '' + n [ A, [ A, A_u]] =0.
$$
Solving for $B_u ''$ and putting $s = m-n$,
$$
B_u '' = n \gamma u^{m-1} A_u - n u^{n-1}
[ A, [ A, A_u ]].
$$
Substituting $A$ from (3.14) and its derivative
$A_u = X_2 + (n+1) u^n X_3$ into $B_u ''$ from above,
we have
$$
B_u '' = n \gamma u^{m-1} X_2 + n (n+1) \gamma u^{n+m-1} X_3
- n u^{n-1} [ X_1 + u X_2 + u^{n+1} X_3, [ X_1, X_2]].
\eqno(3.17)
$$
Suppose at this point that $X_1$ and $X_2$ do not commute
with each other, then a new vector field can be 
defined as
$$
X_7 = [ X_1, X_2 ].
\eqno(3.18)
$$
Setting $X= X_3$, $Y=X_1$ and $Z= X_2$ in
the Jacobi identity $[X, [Y,Z]]+[Y, [Z,X]] +[Z, [X,Y]]=0$
gives
$$
[X_3, [X_1, X_2]] + [X_1, [X_2, X_3]]+ [X_2, [X_3, X_1]] =0.
\eqno(3.19)
$$
However, using (3.16), the last two terms in (3.19)
are zero, hence (3.19) implies that
$$
[ X_3, X_7 ] =0.
\eqno(3.20)
$$
Consequently, $B_u ''$ reduces to the form
$$
B_u '' = n \gamma u^{m-1} X_2 + n (n+1) \gamma u^{n+m-1} X_3
- n u^{n-1} X_5 - n u^n X_6,
\eqno(3.21)
$$
Two new commutators have been introduced
to write (3.21) defined as
$$
[ X_1, X_7 ] = X_5,
\qquad
[ X_2, X_7 ] = X_6.
\eqno(3.22)
$$
Using (3.22) in the Jacobi identity, the following brackets
result
$$
[ X_2, X_5 ] = [ X_1, X_6 ],
\qquad
[ X_3, X_5 ] =0.
\eqno(3.23)
$$
Finally, integrating $B_u ''$ with respect to $u$ yields
an expression for $B''$,
$$
B'' = \frac{n}{m} \gamma u^m X_2 + \frac{n(n+1)}{n+m} 
\gamma u^{m+n} X_3 - u^n X_5 - \frac{n}{n+1} X_6 + X_4.
\eqno(3.24)
$$
Only one term in (3.15) remains to be satisfied,
namely $[ A, B'' ] =0$. Thus substituting $A$ and
$B''$ into this bracket and using linearity to expand
out, we have
$$
[ X_1 + u X_2 + u^{n+1} X_3, \frac{n}{m} \gamma u^m X_2
+ \frac{n (n+1)}{m+n} \gamma u^{m+n} X_3 - u^n X_5
- \frac{n}{n+1} u^{n+1} X_6 + X_4 ]
$$
$$
= \frac{n}{m} \gamma u^m [ X_1, X_2 ] - u^n [ X_1, X_5]
- \frac{n}{n+1} u^{n+1} [ X_2, X_5] + [ X_1, X_4]
- u^{n+1} [ X_2, X_5]
$$
$$
- \frac{n}{n+1} u^{n+2} [ X_2, X_6] + u [ X_2, X_4]
- \frac{n}{n+1} u^{2n+2} [ X_3, X_6] + u^{n+1} 
[ X_3, X_4].
$$
Therefore, the vector fields must be interrelated in such a way that
the following holds among the coefficients of each power of $u$,
$$
[ X_1, X_4] + u [ X_2, X_4] + \frac{n}{m} \gamma u^m [ X_1, X_2]
- u^n [ X_1, X_5] + u^{n+1} (- \frac{2n+1}{n+1} [ X_2, X_5]
+ [ X_3, X_4] )
$$
$$
- \frac{n}{n+1} u^{n+2} [ X_2, X_6] - \frac{n}{n+1} u^{2n+2}
[ X_3, X_6] =0.
\eqno(3.25)
$$
This completes the proof. 

\begin{theorem} 
There exist nontrivial algebras for the
$X_i$ specified by (3.16), (3.18), (3.20), (3.23) and the
coefficients of powers of $u$ in (3.25), which depend on the
relative values of $m$ and $n$.
\end{theorem}

{\em Proof:}
It is required
to equate the independent powers of $u$ equal to zero.
This has to be done on a case by case basis by putting individual
restrictions on $m$ and $n$, and not all cases are given. 

$(i)$ Suppose none of the powers of $u$ in
(3.25) are equal, hence $n \neq m \neq 1,0$. 
Equating each power of $u$ to zero gives the following
algebra
$$
X_7 = [ X_1, X_2]=0,  \quad
[ X_1, X_5] =0,  \quad
[X_2, X_6]=0,  \quad
\frac{2n+1}{n+1} [X_2, X_5]= [X_3, X_4],
$$
$$
[X_3, X_6]=0,
\qquad
[X_2, X_4]=0, 
\qquad
[X_1, X_4]=0.
$$
At this point, $X_1$ and $X_2$ have be required to commute,
since $X_7=0$ must hold. However, from (3.22), it follows 
that $X_5 =X_6=0$. Moreover, $[ X_1, X_3]=0$ implies
that $X_1$ and $X_3$ differ by a constant, hence $X_2$
and $X_3$ also differ by a constant. Finally,
$[X_1, X_4 ]=0$ implies that $X_1$ and $X_4$ differ
by a constant. Therefore, we can put
$$
X_1 = \kappa X,
\qquad
X_2 = \sigma X,
\qquad
X_3 =X,
\qquad
X_4 = \alpha X.
\eqno(3.26)
$$
Substituting these results into $A$ and $B$, they
take the form
$$
\begin{array}{c}
A= ( \kappa + \sigma u + u^{n+1}) X,  \\
  \\
B =- ( \sigma + (n+1) u^n ) q X + \frac{1}{2} (n+1) p^2 X
+ \dss \frac{n}{m} \gamma \sigma u^m X  
+ \dss \frac{n(n+1)}{m+n}  
\gamma u^{m+n} X + \alpha X.  \\
\end{array}
\eqno(3.27)
$$

$(ii)$ Suppose $n=1$ and $m \neq 1,2,3,4$.
Then the same algebra as (3.26) results and $A$ and $B$
are given by (3.27) with $n$ set equal to one.

$(iii)$ Suppose now that $n=m \neq 0,1$, then
prolongation equation (3.25) reduces to
$$
[X_1, X_4] + u [ X_2, X_4] + u^n ( \gamma X_7 - [ X_1, X_5] )
+ u^{n+1} ( - [ X_2, X_5] - \frac{n}{n+1} [X_1, X_6]
$$
$$
+ [ X_3, X_4] ) - \frac{n}{n+1} u^{n+2} [ X_2, X_6]
- \frac{n}{n+1} u^{2n+2} [ X_3, X_6] =0.
$$
This equation is satisfied provided that the
following brackets hold,
$$
[ X_3, X_6]=0,
\quad
[ X_2, X_6]=0,
\quad
\frac{2n+1}{n+1} [ X_2, X_5] = [X_3, X_4],
$$
$$
\gamma X_7= [ X_1, X_5], 
\quad
[ X_2, X_4]=0,
\quad
[ X_1, X_4]=0,
\eqno(3.28)
$$
in addition to the brackets given in (3.20), (3.22), 
and (3.23). This algebra has a simpler three-element 
realization which satisfies all the commutation relations 
provided that
$$
X_3=0,
\quad
X_4=0,
\quad
X_5 = \gamma X_2,
\quad
X_6=X_2.
\eqno(3.29)
$$
The nonzero commutation relations are given by
$$
[ X_1, X_2] = X_7, \quad
[ X_2, X_7] = X_2, \quad
[ X_1, X_7] =- \gamma X_2.
\eqno(3.30)
$$
The algebra closes and a finite three-element algebra results.

$(iv)$ Suppose that $m=n+1 \neq 0,1$, then prolongation equation
(3.25) implies the algebra
$$
[X_1, X_4] =0,
\quad
[ X_2, X_4]=0,
\quad
\gamma \frac{n}{n+1} [ X_1, X_2]
- \frac{2n+1}{n+1} [ X_2, X_5] +[X_3, X_4]=0,
$$
$$
[X_1, X_5]=0, 
\quad
[X_2, X_6]=0,
\quad
[X_3, X_6]=0.
$$
Recalling that (3.23) must be satisfied, a three element
algebra results if we take
$$
X_2 = X_3,
\quad
X_4=0,
\quad
X_5 =- \frac{n \gamma}{2n+1} X_1,
\quad
X_6= \frac{n \gamma}{2n+1} X_2.
\eqno(3.31)
$$
There is a closed algebra in this case with three nontrivial brackets,
$$
[ X_1, X_2]= X_7,
\qquad
[ X_1, X_7] = - \frac{n \gamma}{2n+1} X_1,
\qquad
[ X_2, X_7] = \frac{n \gamma}{2n+1} X_2.
\eqno(3.32)
$$

$(v)$ The linear case $m=n=1$
generates the following bracket relations
$$
[ X_1, X_4] =[ X_2, X_6]= [ X_3, X_6]=0,
\quad
\gamma X_7 +[ X_2, X_4] - [ X_1, X_5]=0,
$$
$$
[ X_3, X_4] - [ X_2, X_5] - \frac{1}{2}
[ X_1, X_6]=0.
\eqno(3.33)
$$

$(vi)$ The case $m=2$, $n=1$ corresponds to the
classical KdV equation and the brackets must satisfy
$$
[X_3, X_6]=0,
\quad
[ X_2, X_6]=0,
\quad
\frac{1}{2} \gamma X_7 = \frac{3}{2} [ X_2, X_5] - [ X_3, X_4],
$$
$$
[ X_2, X_4] -[ X_1, X_5]=0,
\quad
[ X_1, X_4]=0.
\eqno(3.34)
$$
Since (3.23) must be satisfied, this system is
satisfied if we put
$$
X_3=X_4=0,
\qquad
X_5= - \frac{\gamma}{3} X_1,
\qquad
X_6= \frac{\gamma}{3} X_2.
\eqno(3.35)
$$
There are three nontrivial commutators which take the form
$$
[ X_1, X_2]= X_7,
\qquad
[ X_1, X_7] =- \frac{\gamma}{3} X_1,
\qquad
[ X_2, X_7] =\frac{\gamma}{3} X_2.
\eqno(3.36)
$$
This completes the proof.

\subsection{Conservation Laws.}

One way in which conservation laws can be associated
with this equation is that they correspond to the
existence of exact two-forms contained in the ring
of the forms $\alpha_i$ given by (3.1). Let us
suppose we can find a set of functions $g_i (x,t,u,p,q)$
such that the two-form
$$
\vartheta = g_1 \alpha_1 + g_2 \alpha_2 + g_3 \alpha_3
\eqno(3.37)
$$
satisfies the condition for exactness, $d \vartheta =0$.
This is the integrability condition for the existence of 
a one-form, $\omega$, such that
$$
\vartheta = d \omega.
\eqno(3.38)
$$
Conversely, (3.38) implies that $d \vartheta =0$
by the usual identity for double exterior derivatives.
Differentiation of (3.37) and substituting (3.2) yields
$$
d \vartheta = ( d g_1 + g_3 \gamma \frac{s}{n} p u^{s-n} \,
dx )  \wedge \alpha_1 + (  d g_2
+ (g_1  + g_3 \gamma p u^s) \, dx) \wedge \alpha_2
+ ( d g_3 - g_2 \, dx) \wedge \alpha_3.
$$
Therefore $d \vartheta \in I$,
and this clearly vanishes $mod \quad \tilde{p}^* (I)$.

As an example of a form $\vartheta$ with the structure (3.37)
corresponding to equation (3.6), consider the
one-form $\vartheta$ which is given in terms of the
$\alpha_i$ in (3.1) with $g_1= - \gamma u^s$, $g_2=0$
and $g_3=1$ as
$$
\vartheta =- \gamma u^s \, \alpha_1 + \alpha_3,
\eqno(3.39)
$$
where $s=m-n$. Exterior differentiation
$d \vartheta$ gives 
$$
d \vartheta = \gamma s u^{s-1} p \, du \wedge dx \wedge dt
+ \gamma u^s \, dp \wedge dx \wedge dt
- \gamma sp u^{s-1} \, du \wedge dx \wedge dt
- \gamma u^s \, dp \wedge dx \wedge dt =0.
$$
Thus the exterior derivative of (3.39) does vanish.
Substituting $\alpha_1$ and $\alpha_2$ into (3.39),
an explicit form for $\vartheta$ is obtained
$$
\vartheta =- \gamma \frac{n}{m} \, d (u^m) \wedge dt + du \wedge dx
- dq \wedge dt.
$$
If the one-form $\omega$ is defined to be
$$
\omega =- (\gamma \frac{n}{m} u^m \,  + q) \, dt + u \, dx,
\eqno(3.40)
$$
Then it is easy to verify by differentiation that
$\vartheta= d \omega$. The associated conservation law
results from an application of Stokes theorem, which
is written as
$$
\oint_{M_1} \omega = \int_{M_2} \, d \omega.
\eqno(3.41)
$$
This has been written for any simply-connected, two-dimensional
manifold $M_2$ with closed one-dimensional boundary $M_1$. 
The equation implies that $\omega$ and $d \omega$ are
to be evaluated on their respective manifolds.

Returning to $\omega$ once more, we can of course add to $\omega$
any exact one-form $d v$, where $v$ is an arbitrary scalar function.
Thus, $\omega$ can also be taken to be
$$
\omega = dv - (\gamma \frac{n}{m} u^m \,  + q) \, dt + u \, dx,
\eqno(3.42)
$$
such that $\vartheta = d \omega$. Now $v$ may be regarded
simply as a coordinate in an extended six-dimensional
space of variables $\{ x,t,u,p,q \}$, and the one-form
$\omega$ may be included with the original set of forms.
Since $d \omega$ is known to be in the ring of the
original set, the new set of forms remains a closed ideal.

%\begin{center}
\section{Prolongation of a Differential System Related to the
Camassa-Holm Equation.}
%\end{center}

\subsection{ Exterior System and Associated Partial Differential Equation.}

A differential system will be introduced which is
related to several equations which are of interest
in mathematical physics at the moment.
In particular, the Camassa-Holm and Degasperis-Procesi
equations are to be included in this group.
Define the following system of two forms
$$
\alpha_1 = du \wedge dt -p \, dx \wedge dt,
$$
$$
\alpha_2 = dp \wedge dt - q \, dx \wedge dt,
\eqno(4.1)
$$
$$
\alpha_3 =- du \wedge dx + dq \wedge dx + u \, du \wedge dt
- u \, dq \wedge dt + \beta (u-q) \, du \wedge dt,
$$
where $\beta$ in (4.1) is a real, non-zero constant. 
Exterior differentiation of the system of $\alpha_j$ in (4.1)
yields
$$
d \alpha_1 =- dp \wedge dx \wedge dt = dx \wedge \alpha_2,
$$
$$
d \alpha_2 = \frac{1}{u} \, dx \wedge
( - \alpha_3 + u (( 1 + \beta) u -q) \alpha_1),
\eqno(4.2)
$$
$$
d \alpha_3 = (1- \beta) \, dq \wedge du \wedge dt =
(1- \beta) [ dq \wedge \alpha_1 + p \, dt \wedge \alpha_3 - 
p \, dx \wedge \ \alpha_1 ].
$$
Clearly all of the $d \alpha_i$ vanish modulo the set
of $\alpha_j$ in (4.1). Therefore, the ideal
$I = \{ \omega | \omega = \sum_{i=1}^3 \, \sigma_i \wedge \alpha_i 
: \sigma_i \in \Lambda_p (M), p=0,1,2, \cdots \}$ is closed, $d I \subset I$,
and the system is integrable.

On the transversal manifold, it is determined that
$$
0 = \alpha_1 |_S = S^* \alpha_1 = (u_x -p) \, dx \wedge dt,
$$
$$
0 = \alpha_2 |_S = S^* \alpha_2=  (p_x -q) \, dx \wedge dt,
\eqno(4.3)
$$
$$
0 = \alpha_3 |_S = S^* \alpha_3 = ( u_t - q_t + u ( u_x - q_x)
+ \beta (u-q) u_x ) \, dx \wedge dt.
$$
Thus sectioning the differential system (4.1)
generates the following set of equations
$$
p= u_x,
\qquad
q= p_x,
\qquad
(u-q )_t + u(u-q)_x + \beta (u-q) u_x =0.
\eqno(4.4)
$$
The first two equations here imply that
$q=u_{xx}$. Using this in the third
equation of (4.4), the following partial differential
equation results
$$
( u - u_{xx} )_t + u ( u- u_{xx})_x +
\beta ( u - u_{xx}) u_x =0.
$$
The reason for the interest in this equation is that it  
produces important, relevant equations which are of 
current interest when $\beta$ is picked appropriately.
To write this in a concise form,
it is usual to introduce the variable $ \rho = u- u_{xx}$,
which gives
$$
\rho_t + \rho_x u + \beta \rho u_x =0.
\eqno(4.5)
$$
Setting $\beta=2$ in (4.5), the Camassa-Holm equation
results,
$$
\rho_t + \rho_x u + 2 \rho u_x =0.
\eqno(4.6)
$$
For the case in which $\beta =3$, equation (4.5) takes the
form of the Degasperis-Procesi equation,
$$
\rho_t + \rho_x u + 3 \rho u_x =0.
\eqno(4.7)
$$
The results which are obtained below will have
implications for these two equations.

\subsection{ Prolongation Equations.}

Upon substituting differential system (4.1) 
into (3.8), the prolongation relation takes the form,
$$
\frac{\partial A}{\partial t} \, dt \wedge dx
+ \frac{\partial A}{\partial u} \, du \wedge dx
+ \frac{\partial A}{\partial p} \, dp \wedge dx
+ \frac{\partial A}{\partial q} \, dq \wedge dx
$$
$$
+ \frac{\partial B}{\partial x} \, dx \wedge dt
+ \frac{\partial B}{\partial u} \, du \wedge dt
+ \frac{\partial B}{\partial p} \, dp \wedge dt
+ \frac{\partial B}{\partial q} \, dq \wedge dt
+ [A, B] \, dx \wedge dt
\eqno(4.8)
$$
$$
= \lambda_1 ( du \wedge dt -p \, dx \wedge dt)
+ \lambda_2 (dp \wedge dt - q \, dx \wedge dt)
$$
$$
+ \lambda_3 ( - du \wedge dx + dq \wedge dx + u \, du \wedge dt
- u \, dq \wedge dt + \beta (u-q) \, du \wedge dt).
$$
By comparing the coefficients of
the forms on both sides of (4.8), the following
system of prolongation equations is seen
to hold,
$$
\frac{\partial A}{\partial u} =- \lambda_3 ,
\qquad
\frac{\partial A}{\partial p}=0,
\qquad
\frac{\partial A}{\partial q} = \lambda_3,
$$
$$
\frac{\partial B}{\partial u} = \lambda_1 + u \lambda_3 + 
\beta (u-q) \lambda_3,
\qquad
\frac{\partial B}{\partial p} = \lambda_2,
\qquad
\frac{\partial B}{\partial q} =-u \lambda_3,
\eqno(4.9)
$$
$$
- \frac{\partial A}{\partial t} + \frac{\partial B}{\partial x}
+ [ A, B] =-p \lambda_1 - q \lambda_2.
$$
It will be seen that prolongation system (4.9) is much more restrictive
than that obtained for the previous case.

\begin{theorem} 
System (4.9) can be reduced to a single
equation which relates the functions $A$ and $B$.
\end{theorem}

{\em Proof:}
The results obtained in (4.9) imply the following
system of equations,
$$
\frac{\partial A}{\partial u} =- \frac{\partial A}{\partial q},
\qquad
\frac{\partial B}{\partial q} =- u \frac{\partial A}{\partial q}
= u \frac{\partial A}{\partial u},
\qquad
\lambda_1 = \frac{\partial B}{\partial u} - u \lambda_3 -
\beta (u-q) \lambda_3.
\eqno(4.10)
$$
Making use of these in the last equation of (4.9),
we obtain
$$
- \frac{\partial A}{\partial t}
+ \frac{\partial B}{\partial x} +[A,B] =
-p ( \frac{\partial B}{\partial u} - u \lambda_3
- \beta (u-q) \lambda_3 ) - q \lambda_2.
\eqno(4.11)
$$
As with the previous equation studied in Section 3,
$A$ and $B$ are taken to be independent of both
$x$ and $t$ for the same reason. Moreover, the
equation $A_p=0$ implies that $A$ is independent of
the variable $p$, therefore,
$$
A= A( u,q,y),
\qquad
B = B (u, p, q,y).
\eqno(4.12)
$$
Integrating the equation $B_q =-u A_q$ with respect to
$q$, it is found that $B$ is related to $A$ by
$$
B = - u A(u,q,y) + B' (u,p,y).
\eqno(4.13)
$$
In (4.13), $B'$ satisfies $B_p '= \lambda_2$,
but is arbitrary otherwise.
Differentiating (4.13) with respect to $u$,
we obtain
$$
B_u =-A(u,q,y)  -u A_u (u,q,y) + B_u ' (u,p,y).
\eqno(4.14)
$$
The first partial differential equation in (4.10)
$$
\frac{\partial A}{\partial u} + 
\frac{\partial A}{\partial q} =0,
\eqno(4.15)
$$
implies that $A$ must be of the form
$$
A (u,q,y) = A( u-q,y).
\eqno(4.16)
$$
Since $\lambda_3 =-A_u$,
we can write $\lambda_1$ from (4.10) as
$$
\lambda_1 = B_u + u A_u + \beta (u-q) A_u =- A - u A_u 
+ B_u' + u A_u + \beta (u-q) A_u.
\eqno(4.17)
$$
The $u A_u$ terms cancel in (4.17) and so
substituting (4.17) and $B$ from
(4.13) into (4.11), we obtain that
$$
[ A, B' ] =-p (-A + B_u' + \beta (u-q) A_u ) -q B_p'.
\eqno(4.18)
$$
This is the required result, and finishes the proof.

It remains to find solutions to equation (4.18), which
contains two unknown functions. One way to do this is to
impose a condition on one of the unknown functions.

\begin{theorem} 
There exists a nontrivial solution to system (4.15) and (4.18)
under the condition $B' =0$.
\end{theorem}

{\em Proof:}
It follows from (4.15) that $A$
must satisfy (4.16).
Putting $B'=0$ into (4.18), it simplifies to
one equation in one unknown,
$$
\beta (u-q) A_u -A =0.
\eqno(4.19)
$$
On account of (4.16), we can introduce the variable
$\xi =u-q$, and (4.19) then becomes an ordinary
differential equation for $A$
$$
\beta \xi A_{\xi} - A =0.
$$
This equation has the nontrivial solution
$$
A (\xi, y) = (u-q)^{\frac{1}{\beta}} X(y),
$$
with the integration constant written as $X(y)$. 
To summarize explicitly, by using (4.13), the solutions for $A$ and $B$ are 
given as
$$
A( u, q,y) = (u-q)^{\frac{1}{\beta}} X(y),
\qquad
B (u,p,q,y) =- u (u-q )^{\frac{1}{\beta}} X(y).
\eqno(4.20)
$$
$\clubsuit$

Many prolongations can be specified by introducing
different conditions on $B'$. One more will be derived.

\begin{theorem}
There exists a nontrivial solution of (4.15) and (4.18)
such that $B' = \frac{1}{2} (p^2 - u^2) X_2 (y)$ 
with $[ A, X_2 ] =0$.
\end{theorem}

{\em Proof:}
Since $B_p' =p X_2$ and $B_u' =-u X_2$, and $[A, B']=0$,
substituting these, equation (4.18) takes the form
$$
\beta ( u-q) A_u - A = (u-q) X_2.
$$
Using (4.16), this reduces to an ordinary differential equation in the variable
$\xi = u-q$, namely,
$$
\beta \xi A_{\xi} - A = \xi X_2.
$$
This equation has the general solution which for
$\beta \neq 1$ is given by
$$
A ( u, q, y) =  (u-q)^{\frac{1}{\beta}} X_1 (y)
+ \frac{u-q}{\beta -1} \, X_2 (y),
$$
and from (4.13), $B$ is given by
$$
B=- u (u-q)^{\frac{1}{\beta}} X_1 (y) - u \frac{u-q}{\beta -1} X_2 (y) 
+ \frac{1}{2} (p^2 - u^2 ) X_2 (y).
$$
Here, $X_1$ and $X_2$ generate a commutative algebra 
such that $[ X_1, X_2] =0$.
$\clubsuit$

\subsection{ Conservation Laws.}

With the $\alpha_i$ given by (4.1), we can define a form
$\vartheta$ for this case as well. To do the calculations
here, the exterior derivatives (4.2) of the forms
(4.1) can be simplified to read
$$
d \alpha_1 =-dp \wedge dx \wedge dt,
\qquad
d \alpha_2 = - dq \wedge dx \wedge dt,
\qquad
d \alpha_3 = (1 - \beta) dq \wedge du \wedge dt.
$$
Consider the one-form $\vartheta$ which is defined
to be
$$
\vartheta =  - (1- \beta ) q \alpha_1 +
( 1- \beta) p \alpha_2 + \alpha_3.
\eqno(4.21)
$$
Thus, the exterior derivative $d \vartheta$ is in ideal $I$ since
$$
d \vartheta = ( - ( 1 - \beta) \, dq + (1- \beta)p
(( 1 + \beta)u -q) \, dx + (1- \beta) \, dq -p \, dx) \wedge \alpha_1
+ ( 1 - \beta) ( - q dx + dp ) \wedge \alpha_2
$$
$$
+ (1- \beta) ( - \frac{p}{u} \, dx + p \, dt ) \wedge \alpha_3.
$$

Calculating the exterior derivative of $\vartheta$
in terms of the basic set of variables of $M$,
it is found to vanish identically as well,
$$
d \vartheta = (1- \beta ) \, dq \wedge du \wedge dt
- (1- \beta) \, dq \wedge du \wedge dt
+ ( 1- \beta) p \, dq \wedge dx \wedge dt
+ (1 - \beta) q \, dp \wedge dx \wedge dt
$$
$$
- ( 1 - \beta) q \, dp \wedge dx \wedge dt
- ( 1 - \beta) p \, dq \wedge dx \wedge dt =0.
$$
In fact, $\vartheta$ can be obtained directly
from the one-form $\omega$ defined by
$$
\omega= (q-u) dx + \frac{1}{2} ( u^2 - 2 u q + 
\beta u^2 + p^2) \, dt.
\eqno(4.22)
$$
Upon differentiating $\omega$ it is found that
$$
d \omega =- du \wedge dx + dq \wedge dx + u \, du \wedge dt
- u \, dq \wedge dt + \beta u \, du \wedge dt
- q \, du \wedge dt + (1- \beta) p \, dp \wedge dt.
\eqno(4.23)
$$
This is precisely the two-form $\vartheta$ given in (4.21). 
The associated conservation law results from an application
of Stokes theorem, which is written in this case
$$
\oint_{M_1} \omega = \int_{M_2} \, d \omega.
\eqno(4.24)
$$
This has been written for any simply-connected,
two-dimensional manifold $M_2$ with closed
one-dimensional boundary $M_1$. The equation implies that
$\omega$ and $d \omega$ are to be evaluated
on their respective manifolds.

Returning to $\omega$ again, we can again add to
$\omega$ any exact one-form $d v$, where $v$ is
an arbitrary scalar function. Thus, $\omega$
can also be taken to be
$$
\omega= dv + (q-u) \, dx + \frac{1}{2}
( u^2 - 2 u q + \beta u^2 + p^2) \, dt,
\eqno(4.25)
$$
such that $\vartheta = d \omega$. As before, $v$ may
be regarded as a coordinate in an
extended six-dimensional space of variables
$\{ x, t, u, p,q,v \}$, and the one-form
$\omega$ may be included with the original set of
forms. Since $d \omega$ is known to be in the
ring of the original set, the new set of forms
remains a closed ideal.

\section{Summary and Conclusions.}

It has been seen that exterior differential systems
have been constructed for some very important classes of
partial differential equation. As well as giving some 
information about the associated integrability of
these equations, it has been shown that the prolongation 
structure of these systems can be studied. This is more
than just of theoretical interest, since B\"acklund transformations
can be constructed based on these results.
The relationship of differential systems to B\"acklund transformations
has been discussed by Estabrook and Wahlquist {\bf [15]}, and the
construction of such transformations has been done for some
three element algebras in {\bf [16]}. Let us show how to use the
results of example $(i)$ in Section 3 to obtain such a result.

Using (2.4), the connection $\tilde{\omega}$ can always be
chosen on $\mathbb R$ with coordinate $y$ and $X = \partial/ \partial y$
$$
\tilde{\omega} = dy - \{ ( \kappa + \sigma u + u^{n+1} ) \, dx
+ (-( \sigma +(n+1) u^n )q + \frac{1}{2} (n+1) p^2 + \frac{n}{m}
\sigma \gamma u^m 
$$
$$
+ \frac{n (n+1)}{m+n} \gamma u^{m+n} + \alpha
) \, dt \} \, X(y),
\quad \kappa, \sigma, \alpha \in \mathbb R.
\eqno(5.1)
$$
Solutions of the system (3.6) determine transversal sections
of the fibre bundle such that, upon substituting $p$ and $q$ from (3.4),
we have
$$
\begin{array}{c}
y_x = \kappa + \sigma u + u^{n+1},   \\
  \\
y_t =- ( \sigma + (n+1) u^n ) (u^n)_{xx} + \frac{1}{2} (n+1) 
((u^n )_x)^2 + \dss \frac{n}{m} \gamma \sigma u^m + \dss \frac{n(n+1)}{m+n} 
\gamma u^{m+n} + \alpha.
\end{array}
\eqno(5.2)
$$
A similar result but for a different algebra was given in {\bf [9]}.
By solving the first of these for $u$, it can be eliminated
in the second equation of (5.2) to yield an equation for $y=y(x,t)$.
For $\sigma=0$, this can be done in closed form, and to make the
presentation more concise we put $\kappa=0$ as well giving
$$
u = (y_x)^{\frac{1}{n+1}}.
\eqno(5.3)
$$
The positive root is taken if the exponent
in (5.3) has an even denominator.
Eliminating $u$ from the second equation in (5.2), 
we have an equation for $y$
$$
y_t + (n+1) (y_x)^{\frac{n}{n+1}} ( (y_x)^{\frac{n}{n+1}})_{xx} - \frac{1}{2} (n+1) (((y_x)^{\frac{n}{n+1}})_x)^2
- \frac{n (n +1)}{m+n} \gamma ( y_x)^{\frac{m+n}{n+1}} - \alpha =0.
\eqno(5.4)
$$
It follows that for $\sigma=\kappa=0$, the potential equation in
terms of $y$ which results is given by (5.4). In effect, a B\"acklund 
transformation has been determined and is expressed by (5.2).
This set of equations transforms the original equation into the 
form of its potential equation (5.4).

\vspace{0.25cm}
\begin{center}
{\bf References.}
\end{center}

\noindent
$[1]$ A. C. Newell, Solitons in Mathematics and Physics, SIAM,
Philadelphia, 1985.  \\
$[2]$ M. J. Ablowitz and P. A. Clarkson, Solitons, Nonlinear Evolution
Equations and Inverse Scattering, Cambridge University Press, 1991.  \\
$[3]$ H. D. Wahlquist and F. B. Estabrook, Prolongation Structures
of Nonlinear Evolution Equations, J. Math. Phys. {\bf 16}, 1-7, (1975).  \\
$[4]$ H. D. Wahlquist and F. B. Estabrook, Prolongation Structures of 
Nonlinear Evolution Equations II, J. Math. Phys. {\bf 17}, 1293-1297 (1976).  \\
$[5]$ F. B. Estabrook, Moving Frames and Prolongation Algebras,
J. Math. Phys. {\bf 23}, 2071-2076 (1982).  \\
$[6]$ W. F. Shadwick, The KdV Prolongation Algebra, J. Math.
Phys., {\bf 21}, 454-461 (1980).  \\
$[7]$ E. van Groesen and E. M. de Jager, 
Mathematical Structures in Continuous Dynamical Systems,
Studies in Math. Phys., vol. 6, North Holland, 1994.   \\
$[8]$ E. M. de Jager and S. Spannenburg, Prolongation structures and
B\"acklund transformations for the matrix Korteweg-de Vries and
Boomeron equation, J. Phys. A: Math. Gen. {\bf 18}, 2177-2189 (1985). \\
$[9]$ P. Bracken, An Exterior Differential System for a Generalized 
Korteweg-de Vries Equation and its Associated Integrability,
Acta Applicandae Mathematicae, {\bf 95}, 223-231 (2007).  \\
$[10]$ T. Tao, Why are solitons stable?, Bulletin of the Amer. Math. Soc.,
{\bf 46}, 1-33, (2009).  \\
$[11]$ P. Bracken, Symmetry Properties of a Generalized Korteweg-de Vries Equation
and Some Explicit Solutions, Int. J. Math. and Math. Sciences, {\bf 2005, 13},
2159-2173 (2005).  \\
$[12]$ E. G. Reyes, Geometric Integrability of the Camassa-Holm
Equation, Lett. Math. Phys., {\bf 59}, 117-131 (2002).  \\
$[13]$ R. Camassa and D. Holm, An Integrable Shallow Water Equation
with Peaked Solitons, Phys. Rev. Letts. {\bf 71}, 1661-1664 (1993).  \\
$[14]$ J. Lenells, Conservation Laws of the Camassa-Holm Equation,
J. Phys. A: Math. Gen. {\bf 38}, 869-880 (2005).  \\
$[15]$ F. B. Estabrook and H. D. Wahlquist, "Prolongation Structures,
Connection Theory and B\"acklund Transformation", in Nonlinear evolution
equations solvable by the spectral transform, Ed. F. Calogero, Pitman, 1978. \\
$[16]$ P. Bracken, The interrelationship of integrable equations,
differential geometry and the geometry of their associated surfaces,
to appear, 2009.  \\
\end{document}